\documentclass[aps,twocolumn,superscriptaddress,showpacs]{revtex4}
\usepackage{graphics,graphicx,dcolumn,bm,fleqn,epic,eepic,float}
\usepackage{amssymb,amsmath,multirow,rotate,color,float}
\usepackage{times}

\begin{document}

\title{How Gossip Propagates}

\author{Pedro G.~Lind}
\affiliation{Institute for Computer Physics, University of Stuttgart
             70569 Stuttgart, Germany}
\affiliation{Centro de F\'{\i}sica Te\'orica e Computacional, 
             Avenida Professor Gama Pinto 2,
             1649-003 Lisbon, Portugal}
\author{Luciano R.~da Silva}
\affiliation{Departamento de F\'{\i}sica, Universidade Federal do Rio 
             Grande do Norte, 59072-970 Natal, Rio Grande do Norte,
             Brazil}
\author{Jos\'e S.~Andrade Jr.}
\affiliation{Departamento de F\'{\i}sica, Universidade Federal do Cear\'a,
             60451-970 Fortaleza, Cear\'a, Brazil}
\author{Hans J.~Herrmann}
\affiliation{Departamento de F\'{\i}sica, Universidade Federal do Cear\'a,
             60451-970 Fortaleza, Cear\'a, Brazil}
\affiliation{IfB, HIF E12, ETH H\"onggerberg, CH-8093 Z\"urich,
             Switzerland}

\date{\today}

\begin{abstract}
We study different mechanisms of gossip propagation on several network
topologies and introduce a new network property, the ``spread
factor'', describing the fraction of neighbors that get to know the
gossip. We postulate that for scale-free networks the spreading time
grows logarithmically with the degree of the victim and prove this
statement for the case of the Apollonian network. Applying our
concepts to real data from an American school survey, we confirm the
logarithmic law and disclose that there exists an ideal number of
acquaintances minimizing the fraction attained by the gossip. 
The similarity between the school survey and scale-free
networks remains even for cases when gossip propagation only occurs
with some probability $q<1$.
The spreading times follow an exponential distribution that can also be
calculated analytically for the Apollonian network. When gossip also
spreads through strangers the situation changes substantially: the
spreading time becomes a constant and there exists no ideal degree of
connectivity anymore.
\end{abstract}


\maketitle


Gossip is so inherent to human nature that it was already worshipped
in the Greek mythology through the many-tongued Pheme. Its impact in
history and in sociology is so large that its origins have been
studied from many points of view~\cite{Allport,Rosnow,Bergmann}.
However, recent insights into the mathematical properties of social
networks~\cite{Barabasi02,Doro03,Watts} and, in particular, those
involving friendships~\cite{Marta}, open up a new way of understanding
how the propagation of gossip depends on the connections between
people.

In the last years many network types have been proposed and
investigated serving to describe phenomena ranging from the Internet,
epidemics, rumor spreading and logistic planning, to earthquake
prediction, neural activity and immunological defenses
\cite{Barabasi02,Doro03,Watts,Vespignani04,Cohen,Zanette}. 
Correspondingly, many properties have been identified to characterize
these networks for various purposes like the degree distribution, 
the shortest path, the cliquishness, the
inbetweenness, etc. The study of gossip propagation, however, requires
still another not yet considered analysis, giving rise to what we call
the ``spread factor'' and the ``spreading time'' which we will
introduce in this Letter and discuss its properties and applications.
\begin{figure}[htb]
\begin{center}
\includegraphics*[width=8.0cm,angle=0]{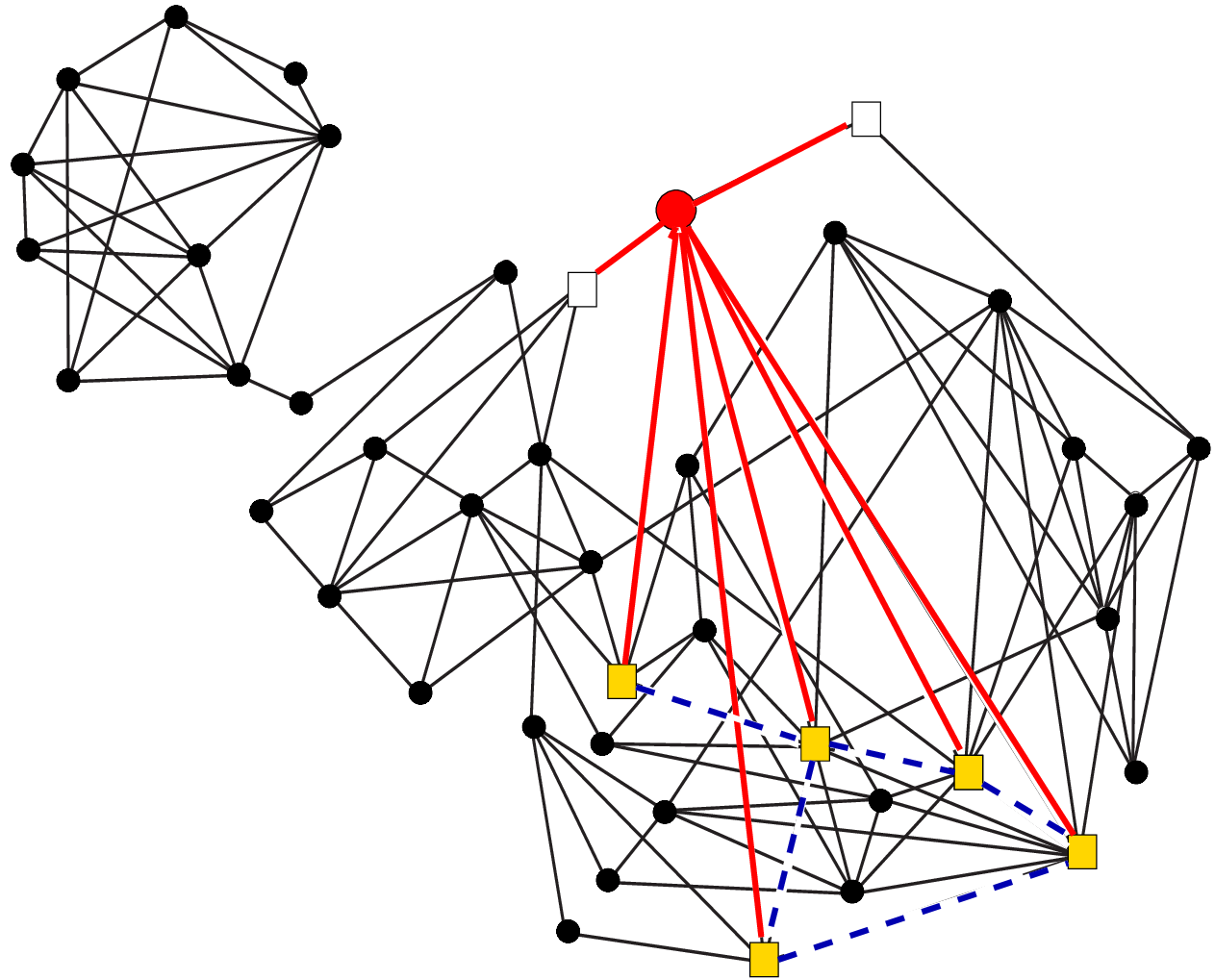}
\end{center}
\caption{\protect 
     (Color online)
     Spreading of a gossip about a victim shown as grey (red) open circle 
     on part of a real school friendship network of Ref.~\cite{schools}.
     If the gossip starts from one of the white squared neighbors, no 
     propagation occurs ($f=0$). If instead, one of the grey (yellow) squared 
     neighbors starts the gossip, in a few time-steps five neighbors will 
     know it, giving $f=5/7$. The gossip spreads over the dashed (blue) lines.
     The clustering coefficient of the victim is $C=10/42$ 
     illustrating the significant difference between $C$ and the 
     spreading factor $f$ (see text). For the white squares $\tau=0$ while
     for the leftmost grey square $\tau=3$.}
\label{fig1}
\end{figure}

As opposed to rumors a gossip always targets the details about
the behavior or private life of a specific person.
Let us consider that individuals are vertices connected by bonds
representing their acquaintance, constituting in this way a
network. Gossip be it truth or falsehood is created at time
$t=0$ about the ``victim'' by the ``originator'' which share a
bond. In the most common case the gossip is only of interest to those
who know the victim personally and we therefore first consider that it
only spreads at each time step from the vertices that know the gossip
to all vertices that are connected to the victim and do not yet know
the gossip. Later we will also consider cases of more famous victims,
like movie stars, about whom gossip can also spread to people they do
not know. Our dynamics is therefore like a burning
algorithm~\cite{burning}, starting at the originator and limited to
sites that are neighbors of the victim. The gossip will spread until
all reachable acquaintances of the victim know it, as illustrated by
the squares connected by dashed lines in Fig.~\ref{fig1} for a real 
friendship network from a recent study~\cite{schools}.

To measure how effectively the gossip attains the acquaintances of the
victim, we define the total number $n_f$ of people who eventually hear
the gossip in a network with $N$ vertices (individuals), and the minimum
time $\tau$ it takes to attain all these vertices which we call the
``spreading time''. In addition, we define the ``spread factor'' as
the fraction of attained vertices, $f=n_f/k$, where $k$ is the degree
of the victim. It is interesting to note that $f$ and the clustering
coefficient $C$ defined in Refs.~\cite{Watts98,Amaral00} are similar
but different because the later one only measures the number of bonds
between neighbors and contains no information about their global
connectivity. 
The quantities $\tau$ and $f$ are averages over all possible 
starting points of the gossip, i.e.~over all neighbors of the 
victim.
Here, we study these quantities on various networks,
namely the Barab\'asi-Albert network~\cite{BA}, the random
graph~\cite{Renyi}, the Apollonian network~\cite{Apoll}, the
small-world lattice~\cite{Watts98} and an empirical friendship network
obtained by interviewing 90118 students from 84
schools~\cite{schools}.
\begin{figure}[t] 
\begin{center}
\includegraphics*[width=8.5cm,angle=0]{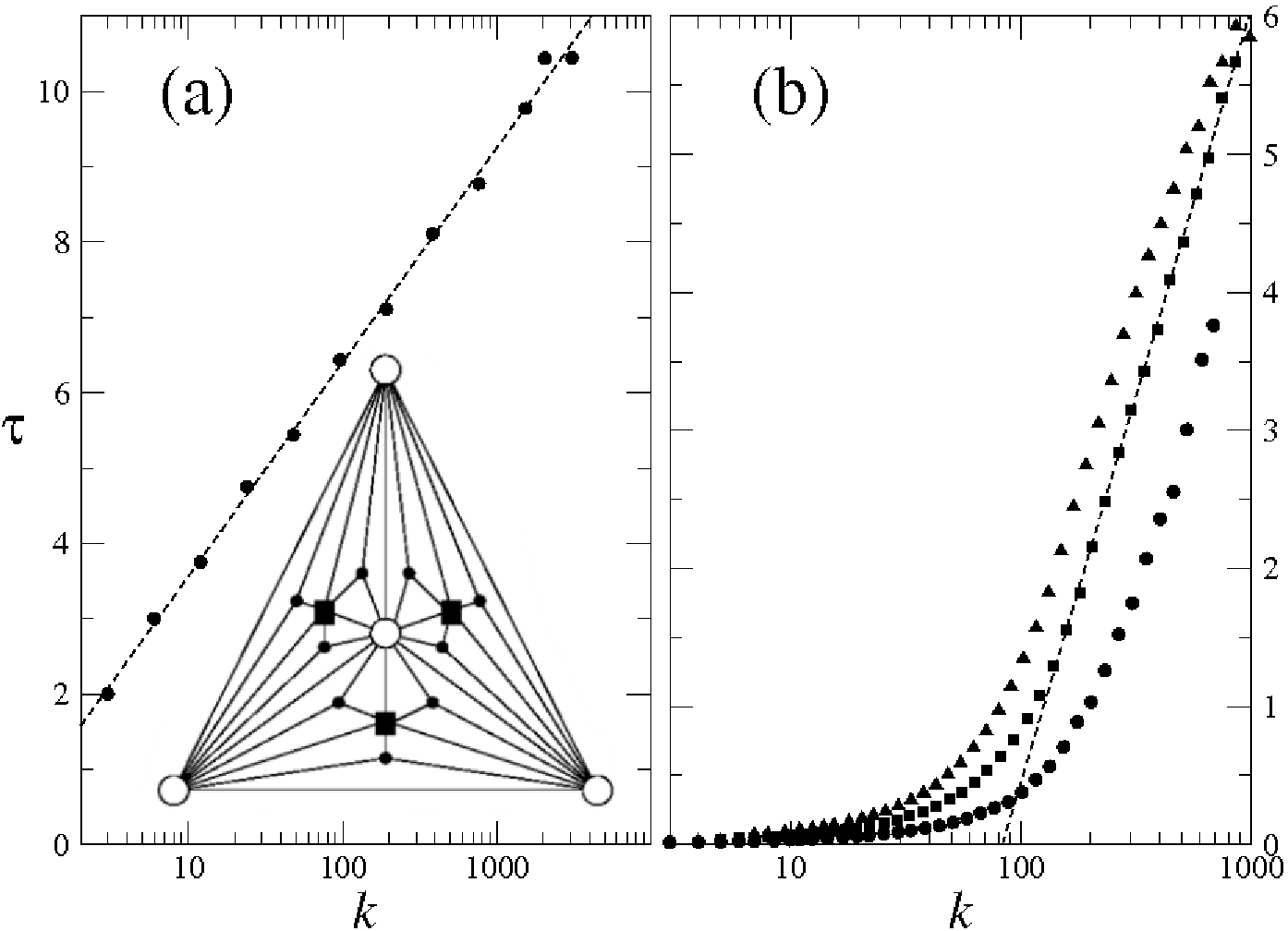}
\end{center}
\caption{\protect 
   Semi-logarithmic plot of the spreading time $\tau$ as a function of
   the degree $k$ for (a) the Apollonian ($n=9$ generations) and (b)
   the Barab\'asi-Albert network with $N=10^4$ nodes for $m=3$
   (circles), $5$ (squares) and $7$ (triangles), where $m$ is the
   number of edges of a new site, and averaged over $100$
   realizations. In the inset of (a) we show a schematic design of the 
   Apollonian lattice for $n=3$ generations. Fitting Eq.~(\ref{eq1}) 
   to these data we have $B = 1.1$ in (a) and $B = 5.6$ for large $k$ 
   in (b).}
\label{fig2}
\end{figure}

In Fig.~\ref{fig2} we see the dependence of the spreading time $\tau$ on the
degree of the victim for two scale-free networks, namely the
Barab\'asi-Albert and the Apollonian ones. In both cases
$\tau$ clearly grows logarithmically, 
\begin{equation}
\tau = A + B {\rm log}k ,
\label{eq1}
\end{equation}
for large $k$. In the case of the Apollonian
network, one can even derive this behavior analytically. As seen from
the inset of Fig.~\ref{fig2}a, in order to communicate between two vertices
of the $n$-th generation, one needs up to $n$ steps, which leads to
$\tau \propto n$. Since, as shown in Ref.~\cite{Apoll}, $k=3 \times
2^{n-1}$, one immediately obtains that $\tau \propto {\rm log}k$. In
fact, we postulate that the spreading time over nearest neighbors is
always asymptotically logarithmic in $k$ for scale-free networks.

\begin{figure}[t] 
\begin{center}
\includegraphics*[width=8.5cm,angle=0]{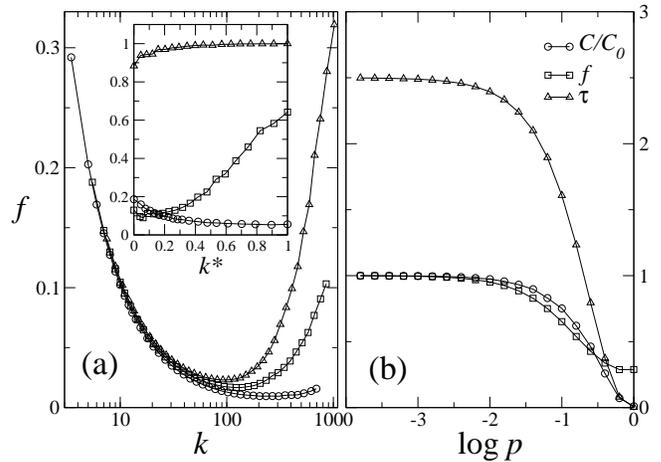}
\end{center}
\caption{\protect 
         (a) Spread factor $f$ for the Barab\'asi-Albert network of
         $N=10^4$ nodes as a function of $k$ 
         for $m = 3$
         (circles), 5 (squares) and 7 (triangles). The inset shows the
         dependence of $f$ on 
         $k^*=(k-k_{min})/(k_{max}-k_{min})$, for the random graph
         with $N=10^3$ sites and $p = 0.02$ (circles), 0.04 (squares)
         and 0.08 (triangles). (b) Spreading time $\tau$, clustering
         coefficient $C/C_0$ and spread factor $f$ as a function of
         the logarithm of the rewiring probability $p$ for the
         small-world lattice of $N=10^4$ sites. $C_0 = {1\over 2}$ is
         the clustering coefficient of a regular lattice. In all cases
         we average over 100 configurations.}
\label{fig3}
\end{figure}

We also see from the inset of Fig.~\ref{fig2}a that for an Apollonian 
network all neighbors of a given victim site can be reached by a gossip 
since they are all connected in a closed path surrounding the victim
site. This means that $f = 1$ which stresses the fact that the spread
factor $f$ is rather different from the clustering coefficient which in
this case is $C=0.828$~\cite{Apoll}. For other networks, the spread factor
is typically less than unity and expresses how many neighbors belong
to the same connected cluster, where the connectivity implies a neighborhood
to the victim. We see in Fig.~\ref{fig3}a the example of the
Barab\'asi-Albert network for different values of $m$ and notice that
there exists a characteristic degree $k_0$ 
for which the relative damage
due to gossip is a minimum. Moreover, larger values of $m$ enhance the
gossip spreading. The inset of Fig.~\ref{fig3}a illustrates the case of
the random graph for different occupation probabilities $p$. We
recognize that below the connectivity threshold $p_c$ of the random 
graph\cite{Renyi}, gossip spreading is very weak, while above $p_c$ 
rapidly the maximum $f = 1$ is attained.

\begin{figure}[htb]
\begin{center}
\includegraphics*[width=8.5cm,angle=0]{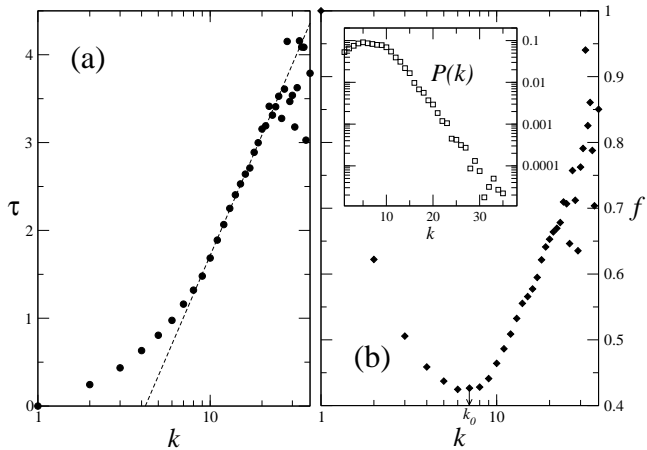}
\end{center}
\caption{\protect 
        Gossip propagation on a real friendship network of American  
        students~\cite{schools} averaged over 84 
        schools. In (a) we show the spreading time
        $\tau$ and in (b) the spread factor $f$, both as a function of
        degree $k$. In the inset of (b) we see the degree distribution
        $P(k)$.}
\label{fig4}
\end{figure}

The linear chain with nearest and second nearest neighbors and a
percentage $p$ of rewired connections, as introduced by Watts and
Strogatz~\cite{Watts98}, produces regular lattices (random graphs) for
small (large) $p$ and small-world networks for intermediate values of
$p$, characterized by large clustering coefficients and small shortest
paths. Figure~\ref{fig3}b illustrates that both the spreading time $\tau$
and the spread factor $f$ 
have the same dependence on $p$ as the clustering coefficient $C$, meaning 
that in small-world societies gossip spreads easily.

Let us now compare these model networks with the real survey data from
Ref.~\cite{schools} averaged over the 84 schools. The data for the
gossip spreading are presented in Fig.~\ref{fig4}. While for small $k$
the spreading time grows linearly, for large $k$ it follows a
logarithmic law like in the case of
scale-free networks. The inset of Fig.~\ref{fig4}b, however, gives 
clear evidence that the school networks are not scale-free. 
As in the case of the
Barab\'asi-Albert networks, we find for the schools a characteristic
degree $k_0$ for which $f$ and therefore the gossip spreading is
smallest. Consequently, there exists an ideal number of friends to
minimize the danger of gossip.

\begin{figure}[htb]
\begin{center}
\includegraphics*[width=8.5cm,angle=0]{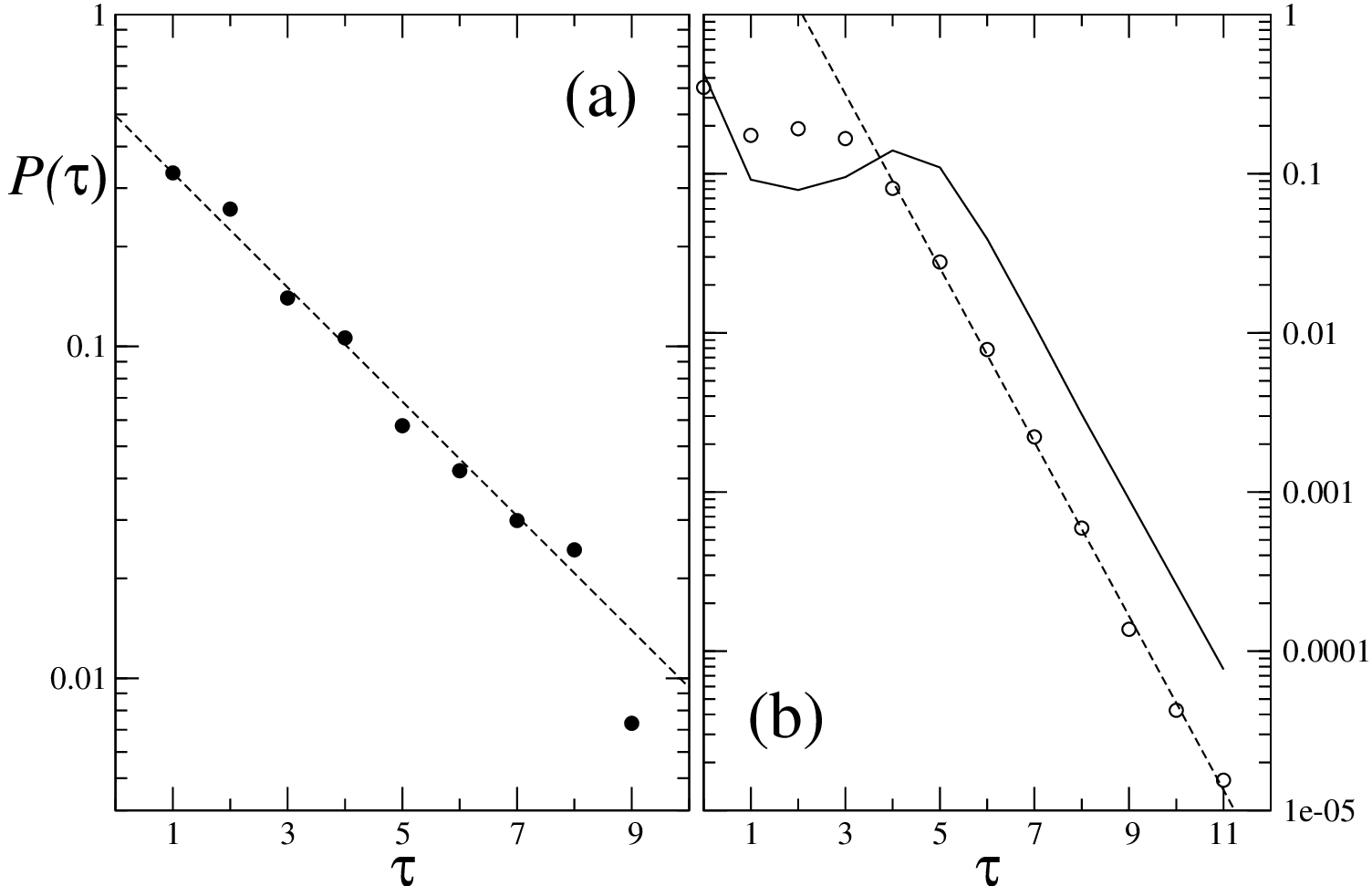}
\end{center}
\caption{\protect 
        Distribution $P(\tau)$ of spreading times $\tau$ for (a) the
        Apollonian network of 8 generations, and (b) the real school 
        network (circles) and the Barab\'asi-Albert network with $m=9$ 
        and $N=1000$ (solid line).}
\label{fig5}
\end{figure}

Another quantity of interest is the distribution $P(\tau)$ of
spreading times. In Fig.~\ref{fig5}a we see that for the Apollonian network
this distribution decays exponentially. This behavior can be
understood if we consider that $P(\tau)d\tau=P(k)dk$ and use Eq.~(\ref{eq1})
together with the degree distribution, $P(k) \propto k^{-\gamma}$, to
obtain 
\begin{equation}
P(\tau) \propto e^{\tau (1-\gamma)/B} ,
\label{eq2}
\end{equation}
for large $k$. The slope in Fig.~\ref{fig5}a is precisely 
$(1-\gamma)/B = -0.17$ using $B$ from Fig.~\ref{fig2}a and $\gamma = 2.58$ 
from Ref.~\cite{Apoll}.
Interestingly, $P(\tau)$ of the school network also follows an
exponential decay for large $\tau$, but with a 3.5 times smaller
characteristic decay time, and has a maximum for small $\tau$, as seen
in Fig.~\ref{fig5}b (circles). 
There we also plot $P(\tau)$ for the Barab\'asi-Albert
network for $m=9$ (solid line), which has a very similar shape but is slightly
shifted to the right, due to the larger minimal number of connections.

\begin{figure}[htb]
\begin{center}
\includegraphics*[width=8.5cm,angle=0]{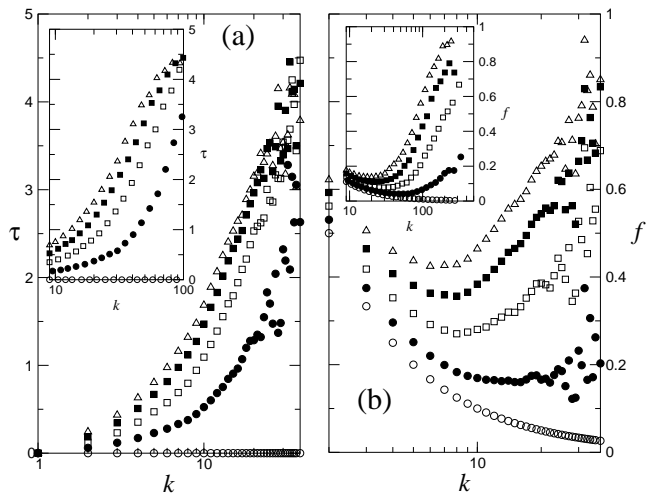}
\end{center}
\caption{\protect
        Gossip propagation on a real friendship network of American  
        students~\cite{schools} averaged over 84 
        schools. In (a) we show the spreading time
        $\tau$ and in (b) the spread factor $f$, both as a function of
        degree $k$. The insets show the same data for the Barab\'asi-Albert
        network with $m=9$ and $N=1000$.
        In all plots one has $q=0$ ($\circ$), $q=0.25$ ($\bullet$), 
        $q=0.5$ ($\square$), $q=0.75$ ($\blacksquare$) and
        $q=1$ ($\triangle$).}
\label{fig6}
\end{figure}

Not everybody likes to gossip, so that it seems realistic to consider
the case in which the transfer from one person to the other only
happens with a given probability $q$. Here we assume that the 
person to which a gossip did not
spread at the first attempt, will never get it. In Fig.~\ref{fig6} we 
see the behavior of $\tau$ and $f$ for different values of $q$ for the
school networks and in the inset for the Barab\'asi-Albert network. 
When the spreading probability $q$ decreases, the minimum in $f$ first 
shifts to larger $k$ and finally disappears.
The asymptotic logarithmic law of $\tau$ for large $k$
seems preserved for all probabilities $q$. As in previous cases,
the Barab\'asi-Albert network has a similar behavior 
as the school friendships. The Apollonian network, 
however, behaves quite differently: $\tau$ first
increases with $q$ and then eventually falls off to zero so that
there exists a special value $q_{max} \approx 0.75$ for which the 
spreading time $\tau$ is maximized.
\begin{figure}[htb]
\begin{center}
\includegraphics*[width=8.5cm,angle=0]{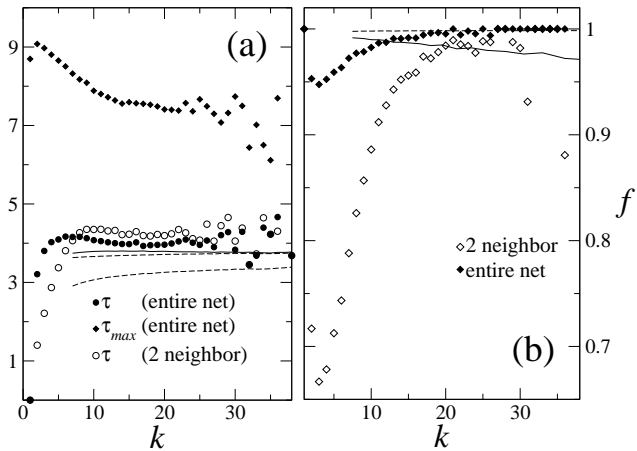}
\end{center}
\caption{\protect 
        Gossip propagation through the first
        two neighborhoods (open symbols) and through the entire network 
        (full symbols) on the school networks of
        Ref.~\cite{schools}. We also show
        results for the Barab\'asi-Albert network with $m=7$ for two
        neighborhoods (solid lines) and the entire network (dashed
        lines). (a) Spreading times as a function of degree $k$.
        For the entire network case, we also show the maximal time
        $\tau_{max}$ for the gossip to reach all students of the
        network (black squares). (b) Spreading factor $f$ as a
        function of degree $k$.
        Here $q=1$.}
\label{fig7}
\end{figure}

As mentioned above, one can also imagine gossip spreading over
strangers, i.e., vertices which are not directly connected to the
victim. We consider the following two cases: ({\it i}) the ``2
neighbor'' case, where besides nearest neighbors also all the
next-nearest neighbors of the victim are involved in gossip spreading,
and ({\it ii}) the ``entire network'' case, where the gossip can spread
over all vertices of the network. This last case could happen to a
public person like a movie star about whom everybody likes to talk. One 
can then ask how long it takes until every attainable person on the network 
will receive the gossip, a time we call $\tau_{max}$.

Astonishingly, in both cases the spreading time $\tau$ and
$\tau_{max}$ on school networks and also on the Barab\'asi-Albert
networks become independent on $k$ as seen in Fig.~\ref{fig7}a. In the ``2
neighbor'' case on the Apollonian network, however, $\tau$ still
increases logarithmically with $k$. The spread factor $f$ looses its
minimum and in the ``entire network'' case even rapidly reaches the
maximum value of unity, as shown in Fig.~\ref{fig7}b. In other words, as
opposed to people living in privacy, famous people are more vulnerable
to gossip and cannot limit the damage by choosing an ideal size for
their friendship network.

In summary, we have established a novel property of networks related to the
spreading of gossip. For real friendship networks stemming from a
survey in 84 American schools, we discovered that, as long as gossip
only spreads among persons knowing the victim, the spreading time
grows logarithmically with the number of friends and that there exists
an ideal number of friends for which the spreading is
minimized. 
The minimum shifts and finally disappears when the probability for gossip
spreading decreases.
The logarithmic law could be derived for Apollonian
networks and we postulate it to be a generic property of scale-free
graphs. The distribution of spreading times of school friendships
is very close to that of the Barab\'asi-Albert network decaying
exponentially for large times as could also be exactly derived for the
Apollonian case. When gossip also spreads among strangers, the non-trivial
dependency on the degree of connectivity disappears. 
Many more details on the issues presented here are being investigated and
will be published in a forthcoming version.
In the future, it would be interesting to study the spreading dynamics 
directly by surveys at different times. 
One of us (LdS) is already making experimental studies on a planetary 
scale which will be published elsewhere.

We profited from discussions with Constantino Tsallis and thank
the Max Planck prize, CAPES, CNPq and FUNCAP for financial support.

\bibliographystyle{prsty}

\end{document}